\documentclass[a4paper,UKenglish]{lipics-v2016}
\usepackage{microtype}

\author[1]{Mikhail Raskin}
\affil[1]{LaBRI, University of Bordeaux \\ \texttt{raskin@mccme.ru}}
\authorrunning{M.\,Raskin}
\Copyright{M.\,Raskin}


\title{A superpolynomial lower bound for the size of non-deterministic complement of an unambiguous automaton
\footnote{
       This work was supported by
       the French National Research Agency (ANR project GraphEn / ANR-15-CE40-0009).
}}
\titlerunning{Superpolynomial lower bound for 1UFA complement}

\EventEditors{Ioannis Chatzigiannakis, Christos Kaklamanis, Daniel Marx, and Don Sannella}
\EventNoEds{4}
\EventLongTitle{45th International Colloquium on Automata, Languages, and Programming (ICALP 2018)}
\EventShortTitle{ICALP 2018}
\EventAcronym{ICALP}
\EventYear{2018}
\EventDate{July 9--13, 2018}
\EventLocation{Prague, Czech Republic}
\EventLogo{eatcs}
\SeriesVolume{80}
\ArticleNo{}

\begin{document}
\maketitle

\begin{abstract}

Unambiguous non-deterministic finite automata have intermediate expressive power
and succinctness between deterministic and non-deter\-mi\-nis\-tic automata.

It has been conjectured \cite{DBLP:conf/dcfs/Colcombet15}
that every unambiguous non-deterministic one-way finite
automaton (1UFA) recognizing some language $L$ can be converted into a 1UFA
recognizing the complement of the original language $L$ with polynomial increase
in the number of states.

We disprove this conjecture by presenting a family of 1UFAs on a single-letter
alphabet such that recognizing the complements of the corresponding languages
requires superpolynomial increase in the number of states even for generic
non-deterministic one-way finite automata.

We also note that both the languages and their complements can be recognized
by sweeping deterministic automata with a linear increase in the number of states.

\end{abstract}

\section{Introduction}

In many areas of computer science, the relationship between deterministic and
non-deter\-mi\-nis\-tic devices is a subject of significant interest.
An intermediate
notion between deterministic and non-deterministic computation devices is the
notion of unambiguous device. Such a device can make non-deterministic choices,
but it is guaranteed that for every input there is at most one accepting
execution trace.

For finite automata it is known that not only non-deterministic automata can be
exponentially more succinct than deterministic automata, but also that
unambiguous automata can be exponentially separated
\cite{DBLP:journals/ijfcs/Leung05}. The paper establishing exponential separation
also defines several automata classes of limited ambiguity and provides exponential
separation between some of them.

Other notions of unambiguity have been considered. Some of them (for example,
structural unambiguity — for a given input word and a given target state
there is at most one way from the initial state) describe a wider class of
automata than unambiguity. Some are more restrictive than simple unambiguity
(for example, strong unambiguity — there is a set of result states, for every
input there is exactly one way to reach a result state, and the result states
can be accepting or rejecting). We do not consider these notions in the
present paper.

We study the problem of representing a complement of a language specified by
a finite automaton. It is easy to see that replacing the set of accepting
states with its complement allows to recognize the complement of a language
specified by a deterministic finite automaton without increasing the number
of states. Complementing a language specified
by a non-deter\-mi\-nis\-tic finite
automaton may require an exponential number of states
\cite{DBLP:journals/tcs/Birget93}.

It has been conjectured \cite{DBLP:conf/dcfs/Colcombet15}
that every unambiguous non-deterministic one-way finite
automaton (1UFA) recognizing some language $L$ can be converted into a 1UFA
recognizing the complement of the original language $L$ with polynomial increase
in the number of states. The best known lower bound was quadratic
\cite{DBLP:journals/iandc/Okhotin12}, while the upper bounds were exponential
\cite{DBLP:conf/dlt/JirasekJS16}. The quadratic lower bound holds even for
the single-letter alphabet.

In the present paper we show a superpolynomial lower bound for the state
complexity of recognizing the complement of a language of a unambiguous
finite automaton 
by a non-deter\-mi\-nis\-tic finite automaton.
The construction
uses only the single-letter alphabet. The family of languages used in the
construction can be recognised both by succinct unambiguous finite automata
and by succinct sweeping deterministic automata. Complementing the language
of a sweeping deterministic finite automata can be done without increasing
space complexity, thus our construction also provides a proof
of superpolynomial state complexity of converting a sweeping deterministic
finite automaton to one-way non-deterministic finite automaton.

\section{Basic definitions and the main result}

In the section we will remind the definitions of deterministic, unambiguous and
non-deterministic finite automata.

\begin{definition}

A $1NFA$ (1-way non-deterministic finite automaton) is defined by an alphabet $\Sigma$,
a set of states $Q$, an initial state $q_0\in Q$, a subset of accepting states $Q_A\subseteq Q$
and the list of permissible transitions $T \subseteq Q\times \Sigma\times Q$.

The size of a $1NFA$ $A$ is the number $|A|$ of states in the definition of the automaton.

A run of a $1NFA$ on a word $w\in\Sigma^*$ is a list of states $q_0, q_1, \ldots q_{|w|}$ such
that all the transitions are permissible, i.e. $\forall i\in 1\ldots |w|: (q_{i-1},w_i,q_i)\in T$.

An accepting run of a $1NFA$ is a run such that the last state is accepting, i.e. $q_{|w|}\in Q_A$.

A $1DFA$ (one-way deterministic finite automaton) is
a $1NFA$ such that for every state and every letter there is at most one permissible
transition, i.e. $\forall q,q_1,q_2\in Q, s\in \Sigma: (q,s,q_1)\in T \wedge (q,s,q_2)\in T \rightarrow q_1=q_2$.

A $1UFA$ (one-way unambiguous non-deterministic finite automaton)
is a $1NFA$ such that for every word there is at most one accepting run.
\end{definition}

We will now define the basic classes of two-way automata for use in the
additional observations about the main construction.
Two-way automata will not be 
used neither in the statement nor in the proof of the main result.

\begin{definition}
A $2NFA$ (2-way non-deterministic finite automaton) is defined by
an alphabet $\Sigma$,
a set of states $Q$, an initial state $q_0\in Q$, a subset of accepting states $Q_A\subseteq Q$
and the list of permissible transitions $T \subseteq Q\times (\Sigma\sqcup\{\vdash,\dashv\})\times
Q\times\{+1,-1,0\}$. We call $\vdash$ and $\dashv$ endpoint markers.

A run of a $2NFA$ on an input word $w\in\Sigma^*$ is
a list of pairs of positions and states,
$(p_0=1,q_0),$ $(p_1,q_1), \ldots,$ $(p_n,q_n)$ such that all transitions are allowed and
the run ends by the only non-moving transition. The exact
conditions are as follows:
        \\ 1) $p_0=1$;
        \\ 2) $\forall i=0..n: 0 \leqslant p_i \leqslant |w|+1$;
        \\ 3) $\forall i=1..n: (q_{i-1},w_{p_{i-1}},q_{i},p_{i}-p_{i-1}) \in T$,
                where we assume $w_0=\vdash$ and $w_{|w|+1}=\dashv$;
        \\ 4) $\forall i=1..n-1: p_i\neq p_{i-1}$;
        \\ 5) $p_n = p_{n-1}$.

An accepting run of a $2NFA$ is a run such that the last state is accepting, i.e. $q_{n}\in Q_A$.

A $2DFA$ is a $2NFA$ such that for every state and every letter there is at most one permissible
        transition, i.e. $\forall q\in Q, s\in \Sigma: \exists! q'\in Q, d\in\{-1,+1,0\}:
        (q,s,q',d)\in T$.

A $swNFA$ (sweeping 2-way deterministic finite automaton) is a $2NFA$ such that every state has
transitions with only one direction of movement (except in case of the endpoint markers and
termination transitions).
Namely, $\neg\exists q,q_1,q_2\in Q, s_1,s_2\in \Sigma: (q,s_1,q_1,+1),(q,s_2,q_2,-1)\in T$.

A $swDFA$ is a $swNFA$ that is also a $2DFA$.
\end{definition}

\begin{definition}

A language $L$ over alphabet $\Sigma$ is an arbitrary subset $L\subseteq \Sigma^*$.

The language recognized by a $1NFA$ $A$ is the set $L(A)$ of all words where there exists an accepting run of $A$.

\end{definition}

\begin{theorem}
There exists a sequence of $1UFA$-s $A_d$ over the single-letter alphabet
such that the minimal $1NFA$ recognising $\overline{L(A_d)}$
has size at least $|A_d|^d$. In other words, complementing a $1UFA$ requires
more than polynomial increase in size regardless of the size of the alphabet,
and the bound holds even if the complement can be representing by $1NFA$.
\end{theorem}

\section{Proof sketch}

We will provide a brief proof outline with a randomized construction.

We consider the single-letter alphabet. We consider only $1UFA$s with the following structure:
there is a set of moduli $m_1,\ldots, m_n$ and some subset of good remainders for each modulus:
$R_i\subseteq \{0,\ldots,m_i-1\}$. The states are $(0)$ and $(i\in1\ldots n, j\in 0\ldots m_i-1)$.
The accepting states are $(i,j\in R_i)$.

A finite automaton over the single-letter alphabet can be illustrated by the directed graph of the 
possible state transitions. The $1UFA$s we have described have graphs with the following structure.

\includegraphics[width=4cm]{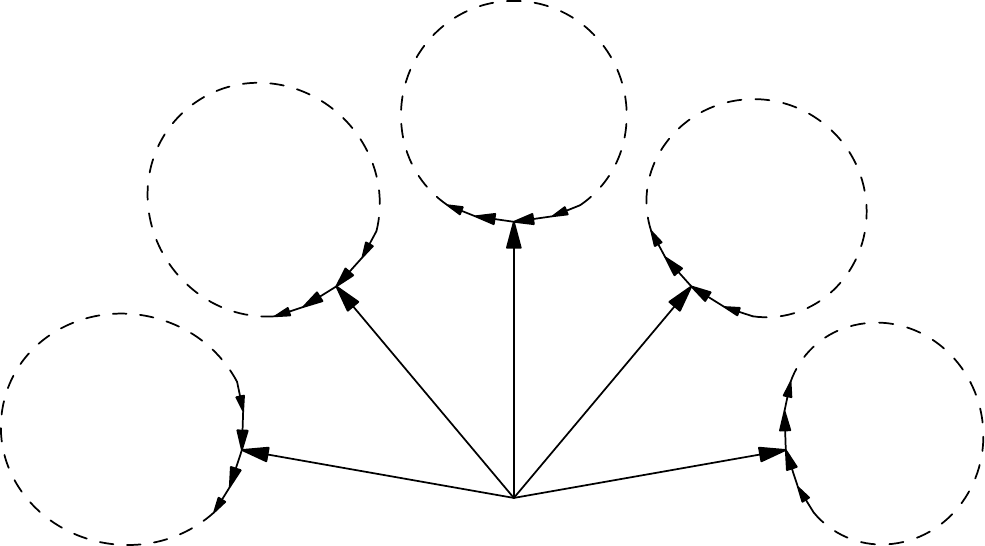}

In other words, on the first step the automaton chooses one of the available
cycles with different lengths, and afterwards it follows the chosen cycle until the end of the word.
(Actually, every $1UFA$ on a single-letter alphabet
can be approximated this way with only a finite number of mistakes,
but we will only use automata that have precisely this form).

We will select a large number $N\gg n$ and some $N$ different primes of roughly the same magnitude,
$P=\{P_1,\ldots,P_N\}$.
For the single-letter alphabet the word is fully described by its
length, and as the moduli are just products of some subsets of $P$, we will only consider
the list of remainders of the lengths in question modulo $P_1, \ldots, P_N$.
By Chinese remainder theorem, every list of remainders is possible.
We will also pick a probability of inclusion $p\in(0;1)$. We will pick each modulus to independently
include each prime with probability $p$, so each modulus will be the product of a different subset of
approximately $pN$ primes. 

Our construction will require that the list
of remainders $(0,\ldots,0)$ is not recognized, but there are many accepted lists with
many zeros in each. At the same time, the automaton must be unambiguous, so every accepted
list must be accepted by unique modulus.

We base our construction of the automaton on an orientation of the complete $n$-vertex
graph. 
Each node will correspond to one modulus.
Thus, we ascribe share $p$ of primes (specifically, the prime factors
of the corresponding modulus)
to each node. The primes ascribed to an edge are the primes ascribed to both
ends of the edge simultaneously (a share $p^2$ of all primes).

\includegraphics[width=5cm]{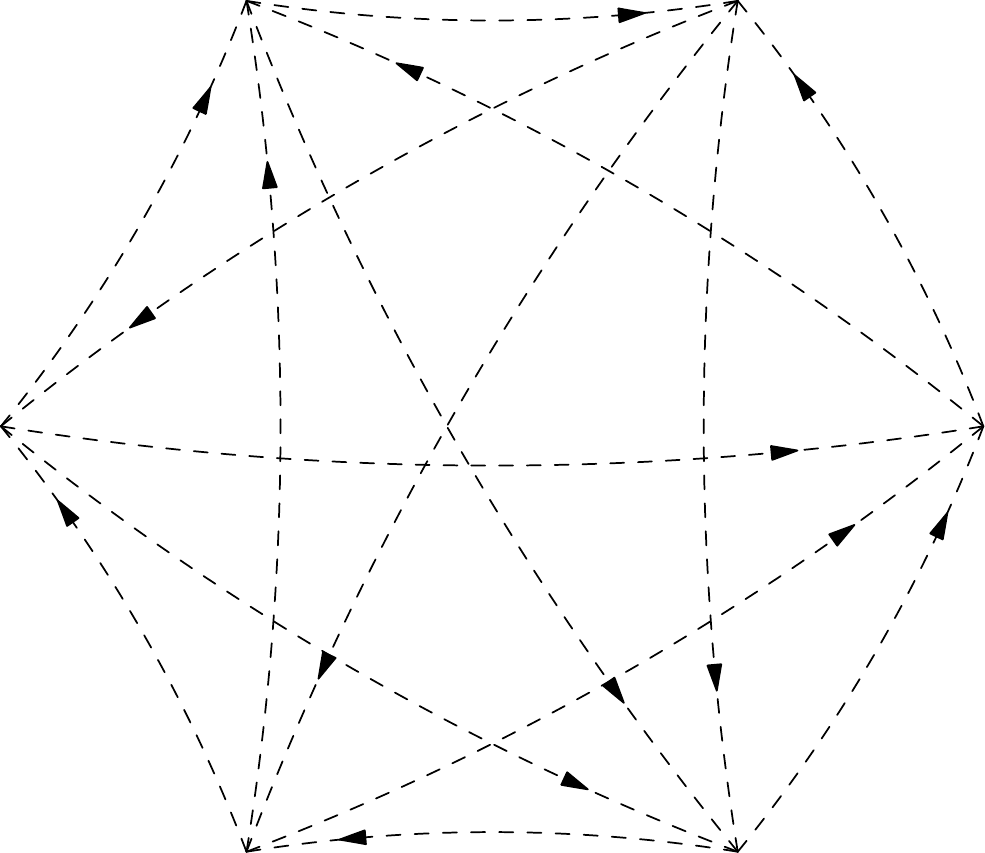}

We will now define the set of accepting states. 
The implementation idea is to treat each remainder as either «no information»
or description of the only vertex (i.e. modulus) that is allowed to accept.
Additionally, checking all the common prime factors for two vertices
and finding only zero remainders
will mean that the source vertex of the edge is not allowed to accept.
For the modulus $m_i$ we will recognize the lists of remainders with the
following properties: all the remainders modulo primes dividing $m_i$ are either $0$ or $i$,
there is at least one remainder equal to $i$, and for every $j$ such that the edge between
$i$ and $j$ in $G$ goes from $i$ to $j$ there is a prime dividing both $m_i$ and $m_j$ such
that the remainder is $i$.

Basically, the remainder $0$ means «no information» and $i$ means «vertex $i$ can recognize».
The vertex $i$ recognizes the situations where it sees only $0$ and $i$
among the remainders modulo its primes, and for every outgoing edge it sees 
at least one $i$ on the edge (an incoming edge can be all-zeros).

Unambiguity of the automaton is easy to verify: if two different moduli $m_i$ and $m_j$
can be used to recognize the same word, without loss of generality we can assume that
the orientation of the edge in $G$ is from $i$ to $j$. Then out of $\approx p^2 N$ primes
shared by $m_i$ and $m_j$ there should be one giving the remainder $i$; but this violates
the condition for acceptance modulo $m_j$.

We can say that two vertices can look at the remainders modulo their 
common primes, and ensure unambiguity: the vertice will not accept
if it sees that some other vertex should, and in case all remainders 
say «no information», edge orientation is used as a tie-breaker.

The size of the $1UFA$ we have constructed is roughly $n\times|P_1|^{pN}$, because we have
$n$ cycles with $pN$ prime factors each.

To ensure that the complement of the language cannot be 
recognized by a small $1NFA$, we will need the orientation to satisfy an
additional condition, which holds for most random orientations of large enough graphs.

The oriented graph will be denoted by $G$. We will want to ensure
that for some $k$ for every choice of $k$ vertices in the graph there is a
vertex outside the chosen subset such that all $k$ edges between the last vertex and the chosen
subset are oriented from the subset towards the last vertex. If $n$ is not large enough relative
to $k$, this is impossible; for large enough $n$ a random orientation will satisfy the condition
with probability close to $1$.
The examples large enough for $k=2$ are already too large to draw
in a small illustration,
so the illustration will use $n=6$. 
We will draw subsets of edges with some properties, with the understanding that for
larger $n$ \textit{all} possible choices have these properties.

Let us estimate the size of a $1NFA$ recognising the complement of this $1UFA$. It has to recognize
all the lengths divisible by the product of all primes in $P$, and almost all of such runs have
to traverse some cycle because of finiteness.
Going around the cycle more times than in this selected run will still produce accepting
runs with the same remainders modulo the primes dividing $C$ and arbitrary remainders modulo
the other ones (by the Chinese Remainder theorem).

Pick any modulus
$m_i$;
if for every $m_j\ne m_i$ there is a common prime factor dividing $m_i$ and $m_j$ but
not $C$, we can build an accepting run  of the $1NFA$ such that its length has
remainder zero modulo every prime dividing $C$ and remainder
$i$ modulo all the other primes in $P$. This run would correspond to a word recognised by the
initial $1UFA$; this contradiction proves that for every modulus $m_i$ there is another
modulus $m_j$ such that the edge in $G$ goes from $i$ to $j$ and every common prime factor
of $m_i$ and $m_j$ divides $C$.

\includegraphics[width=5cm]{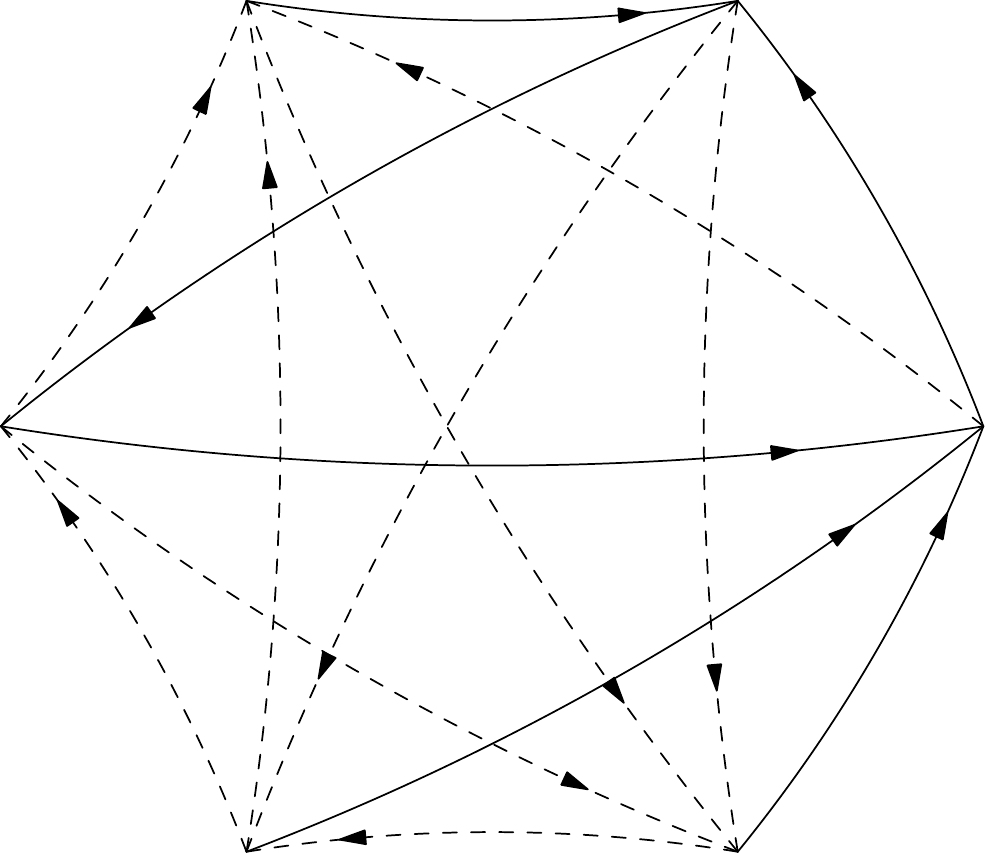}

We can illustrate the choice of primes by drawing solid arrows
when \textit{all} the primes ascribed to the edge divide the 
length of the selected cycle.
To avoid accepting all remainders being zero, we must have at least one
solid outgoing arrow for every vertex.

If for every $k$ vertices in $G$ there is another one with only incoming edges from the $k$
selected ones, we can pick $\frac{k}{2}$ independent edges such that the prime factors
shared by the moduli corresponding to the endpoints of each edge divide $C$.
As long as we have less than $\frac{k}{2}$ edges we can consider all the endpoints, pick
the vertex that doesn't have outgoing edges to the already selected endpoints, and
add the corresponding outgoing edge for this vertex.

\includegraphics[width=5cm]{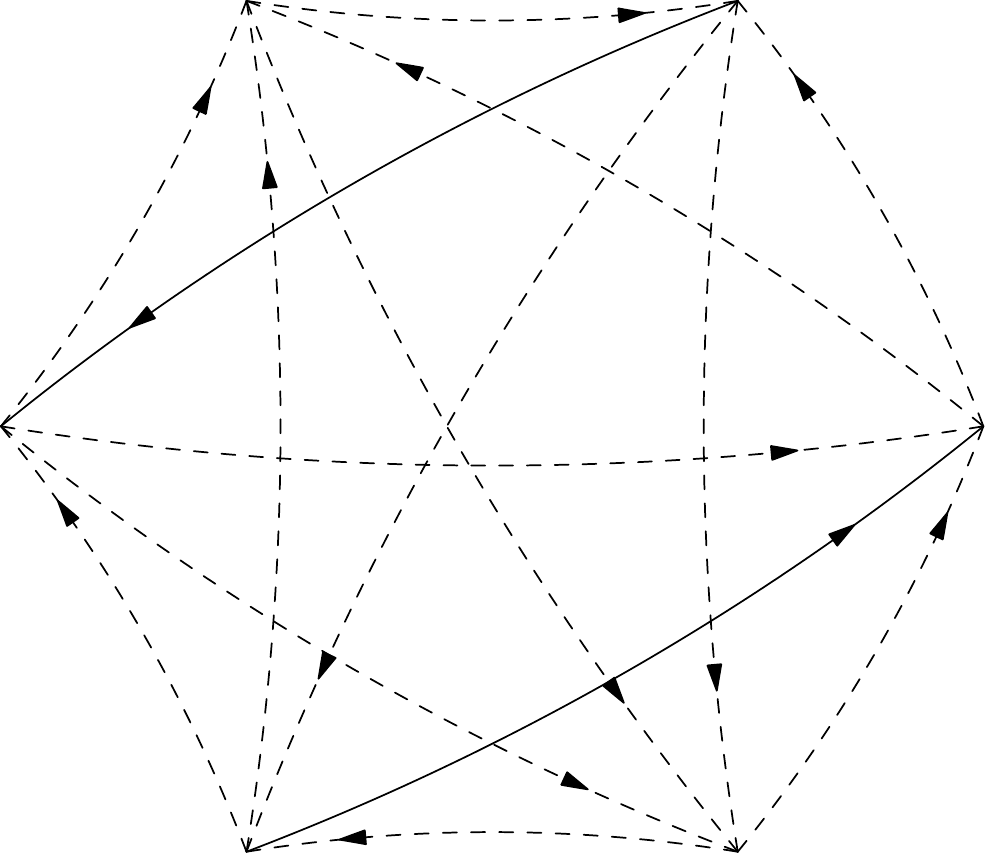}

We can pick just an independent subset of arrows.
Note that our proof doesn't work for the orientation on the picture,
and the graph is too small for a proper orientation;
we just illustrate independence of the edges.

The first edge in such a set forces $C$ to be divisible by $p^2 N$ primes from $P$; each
next edge adds the share $p^2$ of the remaining primes. Therefore, the total number of primes
included will be approximately $1-(1-p^2)^{\frac{k}{2}}$;
for $k>\frac{1}{p^2}$ it will be more than the
half of the entire amount of primes.
So $C$ will be roughly $P_1^{\frac{N}{2}}$, only
a bit less than the size of the initial $1UFA$ to the power $\frac{1}{2p}$.

\section{Graph orientations without small inbound-co\-ve\-ring sets}

In this section we will prove existence
of the orientation of the edges of a complete graph
that is needed for our construction.

\begin{definition}

        Consider a finite complete graph $G$.
        Consider an orientation on its edges.
        The orientation can also be represented as
        an irreflexive anticommutative relation $R\subset V(G)\times V(G)$ on its vertices.
        The relation $R$ holds for a pair of vertices $u$ and $v$ if the edge $(u,v)$ is oriented towards $v$.
        A set of vertices $S\subset V(G)$ is called an \emph{inbound-covering set} if
        every vertex $v$ outside $S$ has at least one edge to $S$ oriented towards $S$.
        We can write this as $\forall v\in V(G)\setminus S: \exists s\in S: R(v,s)$.

\end{definition}

\begin{lemma}

        \label{make-orientation}
        For every positive integer $k$
        there exist a large enough complete graph and an orientation of the graph
        such that the smallest inbound-covering set has size larger than $k$.

\end{lemma}

\begin{proof}

        Consider a uniformly random orientation of a complete graph with $n$ vertices.
        Consider an arbitrary set $S\subset V(G)$.
        For a given vertex $v\in V(G)\setminus S$
        the probability (over the choice of a random orientation)
        of at least one edge between $v$ and $S$ going towards $S$
        is $1-2^{-k}$.
        For a given set $S$ and different vertices $v_1, v_2, \ldots\in V(G)\setminus S$
        existence of an outgoing edge from $v_i$ towards $S$
        is independent for different vertices,
        because the sets of the edges under consideration are disjoint.
        Therefore the probability for a given set $S$ to be inbound-covering
        is equal to $(1-2^{-k})^{(n-k)}$.

        An upper bound can be obtained by approximating the logarithm:
        $\log(1-2^{-k}) < (-2^{-k})$
        and $(1-2^{-k})^{(n-k)} < e^{- 2^{-k}\times (n-k)}$.
        There are at most $n^k$ sets of vertices with size up to $k$,
        and the union (upper) bound for the probability of existence
        of an inbound-covering set of size up to $k$
        is $n^k \exp(-\frac{n-k}{2^k}) = \exp(-\frac{n-k}{2^k} + k \log n)$.

        To prove existence of an orientation
        without inbound-covering sets of size up to $k$
        it is enough to show that this probability is less than $1$.

        \begin{multline}
                \exp(-\frac{n-k}{2^k}+k\log n)<1
                \Leftrightarrow
                -\frac{n-k}{2^k}+k\log n < 0
                \Leftrightarrow
                \frac{n-k}{\log n} > k\times 2^k
                \Leftrightarrow \\
                n > k\times 2^k \times \log n + k
        \end{multline}

        Without loss of generality we can assume that $k$ is at least 8.
        In this case we know that $k>2\log k+\log 3$.
        Now let us pick a large enough $n$, for example $n=3 k^2 \times 2^k$.

        \begin{multline}
                k 2^k \log n + k
                <
                k 2^k \times (k \log 2 + 2 \log k + \log 3) + k
                <
                k2^k \times (2k) + k = \\
                2 k^2 \times 2^k + k
                <
                3 k^2 \times 2^k = n
        \end{multline}

        This inequality proves that a random orientation
        doesn't create any inbound-covering sets of sizes up to $k$,
        and therefore there exists an orientation
        where the smallest inbound-covering set has more than $k$ vertices.

\end{proof}

\section{Building a set of remainders from a graph orientation}

\label{build-remainders}

In this section we will describe how an orientation of the edges of a complete graph
allows to define a set of words.

As we work over the single-letter alphabet,
defining a set of words is the same as defining a set of lengths.
We will define a set of lengths by selecting a large set of distinct primes,
selecting some moduli defined as products of subsets of the selected primes,
and finally selecting some accepted remainders for each selected modulus.
Each modulus will correspond to one vertex of the graph.
We will use the Chinese remainder theorem
to represent the remainders modulo each selected modulus
as a sequence of remainders over the primes in the factorisation of the modulus.

Let us fix an orientation of the edges of a complete graph with $n$ vertices.
Let the relation $R$ represent this orientation.
Let us fix the parameter $b\in\mathbb{N}$. 
This parameter is the inverse of $p$ in the proof overview;
the total number of primes is $b$ times larger than the number of primes
used in any single selected modulus.
Let $N$ be $b^n$,
and let $P=\{P_0,\ldots, P_{N-1}\}$ be a set of distinct primes
within a factor of $1+\frac{1}{N}$ of each other.

\begin{lemma}
For all large enough $N$ it is possible to select $N$ primes 
no larger than $4N^2 \log N$
within a factor of $1+\frac{1}{N}$ of each other.
\end{lemma}

\begin{proof}
We will take the interval of length $3N \log N$
between $3 N^2 \log N$ and $4 N^2 \log N$ that contains the most primes.
By the Prime number theorem there are 
$$\frac{3 N^2 \log N}{2\log N + \log \log N + \log 3} + o(N^2) = \frac{3}{2}N^2 + o(N^2)$$
primes no larger than 
$3 N^2 \log N$
and 
$$\frac{4 N^2 \log N}{2\log N + \log \log N + \log 4} + o(N^2) = 2 N^2 + o(N^2)$$
primes no larger than 
$4 N^2 \log N$. Therefore, there are $\frac{1}{2}N^2$ primes
between $3 N^2 \log N$ and $4 N^2 \log N$.
If we divide this interval into subintervals of length $3N \log N$,
the average subinterval will contain 
$$\frac{1}{2} N^2 \times \frac {3N\log N}{N^2\log N}\times(1+o(1)) = \frac{3}{2}N + o(N)$$
primes, which is enough.
\end{proof}

The modulus $m_i, i\in\{1,\ldots,n\}$ is the product of the primes $P_j$
such that $j$ has $0$ as the $i$-th base-$b$ digit.
Each $m_i$ is the product of $b N$ primes;
the greatest common divisor of $m_i$ and $m_j$ is the product of $b^2 N$ primes,
etc.
Acceptable remainders $r$ modulo $m_i$ are given by the following conditions:
\\ 1) $r$ can only have remainders $0$ and $i$ modulo the primes in $m_i$, i.e.
($\forall q\in P: q | m_i \Rightarrow r \mod q \in \{0,i\}$);
\\ 2) $r$ has at least one non-zero remainder modulo some prime in $m_i$,
i.e. $m_i\not|r$;
\\ 3) for every \emph{outgoing} edge from the vertex $i$ to, say, vertex $j$
there is a prime in $gcd(m_i,m_j)$ such that the remainder is $i$, i.e.
($\forall j: (R(i,j)\Rightarrow\exists q\in P: q|m_i, q|m_j, r\mod q=i$)).

\begin{lemma}\label{unambiguity}

        For every integer $t$ there is at most one selected modulus $m_i$
        such that $t \mod m_i$ is an acceptable remainder modulo $m_i$.

        The remainder $t=0$ is not acceptable modulo any selected modulus.

\end{lemma}

\begin{proof}

        Note that $0$ is not acceptable because it is divisible by every modulus.

        Let $m_i$ and $m_j$ be two distinct selected moduli
        such that
        $t \mod m_i$ is acceptable modulo $m_i$
        and 
        $t \mod m_j$ is acceptable modulo $m_j$.
        Without loss of generality we can assume
        that the edge between $i$ and $j$ goes from $i$ to $j$,
        i.e. $R(i,j)$.
        In this case there is a prime $q$ dividing both $m_i$ and $m_j$
        such that $t\mod q=i$.
        But $q|m_j$ and $(t\mod m_j)\mod q\notin\{0,j\}$.

\end{proof}

\begin{lemma}\label{complement-completeness}

        Assume that the graph orientation has no inbound-covering sets
        of size up to $k$.
        If for some integer $m$
        there are no such $l$ and $i$ such that $m l\mod m_i$
        is an acceptable remainder modulo $m_i$,
        then $m$ has to be divisible by at least
        $N(1-(1-\frac{1}{b}^2)^\frac{k}{2})$
        of the primes in $P$.

\end{lemma}

\begin{proof}

        By the Chinese Remain Theorem,
        for every $i$ we can pick $l$
        such that for every prime $q\in P$ no dividing $m$,
        the remainder $m l \mod q$ is equal to $i$.
        Let us denote such $l$ by $L(m, i)$.

        If for any $m_i$ there is no $m_j$ such that $gcd(m_i,m_j)|m$,
        $m \times L(m,i)$ would be acceptable modulo $m_i$,
        in contradiction with the assumption of the theorem.
        Therefore for every $m_i$
        there is at least one such $m_j$
        that $gcd(m_i,m_j)|m$.
        Let us call edges $i,j$ such that $gcd(m_i,m_j)|m$ \emph{controlled} edges.

        If some vertex $i$ has only incoming controlled edges,
        $m \times L(m,i)$ would still be acceptable modulo $m_i$,
        because the vertex $i$ has to have outgoing edges
        to avoid being an inbound-covering set.
        
        Let us picked $\lceil\frac{k}{2}\rceil$ controlled edges
        that do not share vertices.
        As long as we haven't picked the desired number of edges,
        there are fewer than $k$ endpoints of the selected edges.
        The set of these endpoints cannot be an inbound-covering set,
        so there is a vertex other than these endpoints
        that has only incoming edges from the selected endpoints.
        This vertex has to have an outgoing controlled edge,
        which we can add to the set of the picked controlled edges.

        Now that we have $\lceil\frac{k}{2}\rceil$ independent controlled edges,
        we can count the number of prime factors of $m$
        needed for controlling these edges.
        The prime $P_j$ not being included
        in any of the $\lceil\frac{k}{2}\rceil$ edges
        means that there are $\lceil\frac{k}{2}\rceil$ pairs of positions
        in the base-$b$ representation of $j$,
        and in every pair of positions there should be at least one non-zero digit.
        There are $N(1-(1-\frac{1}{b}^2)^{\lceil\frac{k}{2}\rceil})$ such primes,
        and $m$ is divisible by all of them.

\end{proof}

\section{The main result and its proof}

\begin{theorem}
There exists a sequence of $1UFA$-s $A_d$ over the single-letter alphabet
such that the minimal $1NFA$ recognising $\overline{L(A_d)}$
has size at least $|A_d|^b$. In other words, complementing a $1UFA$ requires
more than polynomial increase in size regardless of the size of the alphabet,
and the bound holds even if the complement can be representing by $1NFA$.

The languages $L(A_d)$ and $\overline{L(A_d)}$ can also be recognised
by a $swDFA$ of size $|A_d|+o(|A_d|)$.
\end{theorem}

\begin{proof}

        Let us fix $d$. Let $b=2d, k=2b^2, n=3k^2\times 2^k, N=b^n$. 

        By lemma \ref{make-orientation}, there exists an orientation
        of complete graph with $n$ vertices
        without inbound-covering sets of size up to $k$.

        Consider the set of moduli and corresponding acceptable remainders
        as built in section \ref{build-remainders}.

        A $1UFA$ can recognize all the lengths 
        having acceptable remainder modulo some modulus
        by guessing the modulus then tracking the remainder.
        This construction requires no more than
        $O(n)\times P_{N-1}^\frac{N}{b}$ states.
        This automaton will be unambiguous by lemma \ref{unambiguity}.
        This is the automaton $A_d$.

        A $swDFA$ can go through the word $n$ times
        calculating the remainder modulo the next modulus each time.
        This construction also requires $O(n)\times P_{N-1}^\frac{N}{b}$ states,
        and can be used to recognize the complement of the language.

        A $1NFA$ recognizing the complement of the language
        has to have a cycle,
        because the complement is infinite.
        There is a cycle that can be traversed
        in a run recognizing some length
        divisible by the product of all the primes in $P$.
        The product of all the primes in $P$
        has remainder zero modulo every modulus $m_i$ in the construction.
        As we can consider the runs of the $1NFA$
        that traverse the cycle multiple times without changing anything else,
        no multiple of the cycle length can be acceptable
        modulo any of the moduli $m_i$.

        By lemma \ref{complement-completeness} this implies
        that the cycle length is divisible
        by at least $N(1-(1-\frac{1}{b}^2)^\frac{k}{2})$ primes from $P$.
        As 
        $(1-\frac{1}{b}^2)^\frac{k}{2} =
        (1-\frac{1}{b})^b < \exp(-\frac{1}{b})^b = \frac{1}{e}$,
        $N(1-(1-\frac{1}{b}^2)^\frac{k}{2}) > N\times(1-\frac{1}{e}) > 0.6N$.
        The total size of the $1NFA$ cannot be less 
        than the size of this cycle,
        which has to be at least $P_0^{0.6N}$.

        We now only need to verify that
        $
        (O(n)\times P_{N-1}^\frac{N}{b})^d < P_0^{0.6N}
        $.
        But indeed, for large enough $d$ we have $P_0 > N\gg n \gg d$ and
        $$
        (O(n)\times P_{N-1}^\frac{N}{b})^d
        <
        ((O(n)\times (1+2\frac{1}{N})P_0)^\frac{N}{2d})^d
        <
        O(n)^{\frac{N}{2}} \exp(\frac{d}{N})P_0^\frac{N}{2}
        <
        P_0^{0.6N}
        $$

        In case of $d$ not large enough, we can replace the automaton with
        the automaton for the smallest large enough $d$.

\end{proof}

\begin{remark}
        The size of $A_d$ is $2^{2^{2^{d^{\Theta(1)}}}}$.
        
        It is sometimes impossible to recognize the complement
        of the language of a $1UFA$ of size $z$ by a $1NFA$ of size less
        than $z^{(\log \log \log z)^{\Theta(1)}}$.
\end{remark}

\begin{proof}
        Let us write down the dependencies between parameters. 
        We know that $b$ is linear in $d$, $k$ is quadratic in $d$, 
        $n$ is $2^{\Theta(d^2)}$, $N$ is $b^n = b^{2^{\Theta(d^2)}} = 
        2^{(\log b)\times 2^{\Theta(d^2)}} = 2^{2^{\Theta(d^2)}}$.
        The primes in $P$ are all $\Theta(N^2\log N)$.
        Then the size of the automaton $A_d$ is 
        $
        \Theta(n\times P_0^\frac{N}{b})=P_0^{\Theta{\frac{N}{b}}}
        =
        (N^2 \log N)^{\Theta{\frac{N}{b}}}
        =
        2^{2^{\Theta(d^2)}\times 2^{2^\Theta(d^2)}}
        =
        2^{2^{2^\Theta(d^2)}}
        =
        2^{2^{2^{d^{\Theta(1)}}}}
        $

The second claim is just a restatement of the same fact.
\end{proof}

\section{Conclusion and further directions}

We have constructed a counterexample to the conjecture
that the complement of a language recognized by a $1UFA$ 
can be recognized by a $1UFA$
with polynomial increase in the number of states.
Moreover, in our example the language and its complement
are easy to recognize by a $swDFA$
with approximately the same number of states,
but the complement requires
superpolynomial number of states in the recognizing $1NFA$
even without the requirement of unambiguity.
The example only uses the single-letter alphabet.

The construction provides a relatively weak kind of superpolynomial growth.
It would be interesting to improve the lower bound.
It seems likely that the number of primes 
used in the construction
could be reduced, making the growth faster.

The question about exponential separation in the case of a general alphabet
remains open.
We hope that disproving the conjecture will inspire new results in this area.

\section{Acknowledgements}

The author is grateful to Gabriele Puppis for numerous useful discussions.

\bibliographystyle{plain}
\bibliography{1ufa-complementing-1nfa-lower-bound}

\begin{thebibliography}{1}

\bibitem{DBLP:journals/tcs/Birget93}
Jean{-}Camille Birget.
\newblock Partial orders on words, minimal elements of regular languages and
  state complexity.
\newblock {\em Theor. Comput. Sci.}, 119(2):267--291, 1993.

\bibitem{DBLP:conf/dcfs/Colcombet15}
Thomas Colcombet.
\newblock Unambiguity in automata theory.
\newblock In Jeffrey Shallit and Alexander Okhotin, editors, {\em Descriptional
  Complexity of Formal Systems - 17th International Workshop, {DCFS} 2015,
  Waterloo, ON, Canada, June 25-27, 2015. Proceedings}, volume 9118 of {\em
  Lecture Notes in Computer Science}, pages 3--18. Springer, 2015.

\bibitem{DBLP:conf/dlt/JirasekJS16}
Jozef~Jir{\'{a}}sek Jr., Galina Jir{\'{a}}skov{\'{a}}, and Juraj Sebej.
\newblock Operations on unambiguous finite automata.
\newblock In Srecko Brlek and Christophe Reutenauer, editors, {\em Developments
  in Language Theory - 20th International Conference, {DLT} 2016,
  Montr{\'{e}}al, Canada, July 25-28, 2016, Proceedings}, volume 9840 of {\em
  Lecture Notes in Computer Science}, pages 243--255. Springer, 2016.

\bibitem{DBLP:journals/ijfcs/Leung05}
Hing Leung.
\newblock Descriptional complexity of nfa of different ambiguity.
\newblock {\em Int. J. Found. Comput. Sci.}, 16(5):975--984, 2005.

\bibitem{DBLP:journals/iandc/Okhotin12}
Alexander Okhotin.
\newblock Unambiguous finite automata over a unary alphabet.
\newblock {\em Inf. Comput.}, 212:15--36, 2012.

\end{thebibliography}

\end{document}